\documentclass[aps,11pt, superscriptaddress, showpacs,showkeys,floatfix]{revtex4}

\usepackage{graphicx}

\begin{document}
\title{Does the Big Rip survive quantization?}

\author{Ed\'esio M. Barboza Jr.\footnote{E-mail: edesio@if.uff.br} and 
 Nivaldo A. Lemos\footnote{E-mail: nivaldo@if.uff.br}}

\affiliation{Departamento de F\'{\i}sica, Universidade Federal Fluminense,\\
Av. Litor\^anea s/n, Boa Viagem  \\
CEP 24210-340, Niter\'oi - RJ, Brazil.}

\date{\today}

\begin{abstract}
 It is known that certain quantum 
cosmological models present quantum behavior for large scale factors. Since quantization can suppress past singularities,  it is natural to inquire 
whether quantum effects can prevent future singularities. To this end,  a Friedmann-Robertson-Walker quantum cosmological model dominated by a 
phantom energy fluid is investigated.  The classical model displays accelerated expansion ending in a Big Rip. The quantization is performed in three  
different ways,  which turn out to lead to the same result, namely  there is a possibility that  quantum gravitational effects could not remove the Big Rip.
\end{abstract}



\maketitle

\section{Introduction}

Ever since there appeared compelling evidence that the Universe is expanding at an increasing rate \cite{Perlmutter} many ideas have been put
 forward to make sense of  this unexpected discovery. 
One of the most popular is the assumption that the  Universe is homogeneously permeated with
 a mysterious dark energy which, although unseen,  dominates the present energy content of the  
Universe (for an up-to-date  review see \cite{Copeland}). Accelerated expansion driven by negative 
pressure is characteristic of a wide variety of  proposals such as cosmological constant \cite{Weinberg}, quintessence \cite{Wetterich},
 $k$-essence \cite{Chiba},
 braneworlds \cite{Dvali}, Chaplygin gas \cite{Kamenshchik},  quintom \cite{Feng}, and 
holography \cite{Cohen}. The generalized Chaplygin gas, described in terms of a complex scalar field, has been first proposed to unify dark matter 
and dark energy in 
\cite{Bilic} and \cite{Bento}.  Quartessence \cite{Makler}, which  also relies on exotic equations of state, and a dark fluid implemented
by a complex scalar field with an exponential potential  \cite{Arbey} are other attempts at  unifying  dark matter and dark energy. 
A striking candidate to dark energy is a so-called  phantom field \cite{Caldwell} whose equation of
 state is $p=w \rho$ with $w <-1$. A recent analysis \cite{Riess}  shows that the possibility that our Universe contains phantom energy 
cannot be ruled out by the present observational data.
Instead of decreasing as the Universe expands, the
 phantom energy density grows without bound and eventually dominates any other form of energy. The scale factor,  as well as the phantom energy density, 
becomes infinite  a finite time from now. This hypothetical catastrophic end of the Universe, with everything torn apart  by the infinite phantom energy 
density, has been christened  a ``Big Rip" \cite{Kamionkowski}.  

Several suggestions have been made either to shun phantom energy or to retain it while  avoiding the Big Rip. The most radical  of such proposals 
 do away with  dark energy altogether either by ascribing  the  observed acceleration of the Universe  to  back-reaction of cosmological
 perturbations \cite{Kolb}
or to a postulated modification of general relativity through the addition of  inverse curvature terms to the Einstein-Hilbert action \cite{Carroll}.
Other models  without dark energy are based on such disparate schemes as  taking the Universe to be an expanding spherical 3-brane \cite{Gogberashvili} 
or imputing the origin of the cosmic 
acceleration to the short range interaction of fundamental particles  \cite{Diez}.
Cosmological evolution in the presence of phantom energy without a  Big Rip seems possible \cite{Piao}, as in the so-called ``hessence" models  \cite{Wei},  in which a
 non-canonical complex scalar field plays the role of dark energy and the equation-of-state parameter can cross the phantom divide $w=-1$ .
It has also been claimed that avoidance of the Big Rip in phantom cosmology can be accomplished  by gravitational back-reaction \cite{Wu}.

Phantom-like behavior can arise  in a dilatonic brane-world scenario with induced gravity \cite{Bouhmadi} or 
be caused by a  non-Hermitian but CPT symmetric Hamiltonian \cite{Andrianov}. At the phenomenological level, in which phantom energy is 
described as a perfect fluid,  thermodynamical considerations \cite{Lima} suggest that phantom energy has negative entropy, 
which appears to defy a sensible physical interpretation.

There have been  studies not only of classical cosmological scenarios motivated by quantum effects and
 containing phantom energy \cite{Nojiri} but also of the influence of quantized fields on the evolution of  classical cosmological models containing a
 Big Rip \cite{Calderon}, in addition to attempts at reformulating semiclassical effects in terms of an effective phantom fluid in loop quantum 
cosmology \cite{Lidsey}.
 There are some examples of quantum cosmological models that present quantum behavior for large scale factors
 \cite{Kowalski}. Since, in many cases, quantization has the power to suppress past singularities, it is reasonable to  inquire whether quantum gravitational 
effects 
might prevent the Big Rip, which is a future singularity.
   
  A series of investigations of quantum effects in  phantom cosmology based on one-loop effective action corrections or back-reaction of quantum fields 
indicate that the
 future singularity may be softened or even suppressed \cite{Odintsov}. It has recently been found that modifications to Friedmann's equation induced 
by loop quantum gravity lead
 to avoidance of  certain future singularities, including the Big Rip \cite{Sami}. Similarly, a quantum phantom cosmology based on the 
Wheeler-DeWitt equation and the representation of phantom energy by a scalar field  has indicated that wave packets tend to disperse in the region that
 corresponds to the Big Rip, thus removing the future singularity \cite{Kiefer}.

Since there are several  independent ways of doing  quantum cosmology, and  it is not at all clear how they are related, one should try to find out  what 
these different formulations  have to tell about future singularities. One widely explored approach to quantum cosmology consists in describing the matter 
content
 by means of fluids, making use of some action principle to describe gravity plus matter  and then performing a canonical quantization.  To our knowledge, 
within  this formalism
no analysis of quantum cosmological models containing phantom energy, with emphasis on  the possible avoidance of the Big Rip by quantum gravitational 
effects, has appeared so far.
 As a preliminary step in this direction, here we 
investigate a quantum cosmological model dominated by phantom energy, 
the latter mimicked
by a perfect fluid.
As will be seen, this quantization scheme  is worth 
pursuing since it 
is exactly soluble  and  produces  interesting results. 
Of course it would be more realistic not to assume that the phantom fluid dominates the dynamics of
 the Universe throughout its history, as we do to make the model tractable. Nevertheless,  this may be considered not too serious a weakness 
since we are interested in the possible influence of quantum effects
 on the very late time dynamics.

The quantization is performed  in three different ways, in order to reach a more satisfactory level of generality and
gain confidence that the results are not a mere artifact of a specious method of quantization.
We find that for a certain class of quantum states  a
 phantom fluid  brings about a Big Rip, just as it does classically.

Here is a summary of the paper. In Section 2 the classical model is formulated on the basis of the canonical formalism of Schutz and the classical 
equations of motion are solved. In the case of a phantom fluid the model displays a Big Rip. In Section 3 a quantization is performed with a particular 
order of noncommuting operators in the Wheeler-DeWitt equation, which requires a special inner product. The  expectation value of the 
scale factor 
grows without bound and becomes infinite at a finite value of cosmic time. In Section 4 the order of noncommuting operators is left arbitrary so that the quantization can be 
carried out with the standard inner product. In Section 5, after reduction of the classical Hamiltonian to that of a free particle with the help of a canonical 
transformation, a new quantization is achieved with the standard inner product. The results of Sections 4 and 5 confirm those of Section 3:  accelerated 
expansion and a Big Rip. Section 6 is devoted to our conclusions.

\section{The Classical Model}

Our starting point is the canonical formalism developed by Schutz \cite{Schutz} to describe gravity plus a perfect fluid. The action is

\begin{equation}
\label{actionSchutz}
{\cal S}= \int_{\cal M} d^4x\, \sqrt{-g}\, R + 2\int_{\partial\cal M}d^3x\, \sqrt{h}\,h_{ab}K^{ab} +
\int_{\cal M} d^4x\, \sqrt{-g}\, p
\,\,\, ,
\end{equation}
\\
where $K^{ab}$ is the extrinsic curvature and $h_{ab}$ is the metric of the boundary $\partial{\cal M}$ of the four-dimensional domain $\cal M$.  Units 
are chosen such that $16\pi G=1$, $c=1$, and $\hbar =1$.
The pressure $p$ is related to the energy density $\rho$ by the equation of state $p=w\rho$ where $w$ is constant. The four-velocity of the fluid is 
expressed in terms of 
five velocity potentials $\epsilon$, $\zeta$, $\beta$, $\theta$, $S$ as

\begin{equation}
\label{four-velocitySchutz}
U_{\nu} = \frac{1}{\mu} (\epsilon_{,\nu} + \zeta \beta_{,\nu} + \theta S_{,\nu})
 \,\,\, ,
\end{equation}
\\
where $\mu$ is the specific enthalpy, and the comma denotes partial derivative. In the case of Friedmann-Robertson-Walker (FRW) models the only 
nonvanishing potentials are $\epsilon$, $\theta$,
 and the specific entropy $S$. The normalization condition $\,U_{\nu} U^{\nu} = -1\,$ determines $\mu$ in terms of the potentials.

The FRW metric is

\begin{equation}
\label{FRWmetric}
 ds^2 = - N^2(t) dt^2 + a^2(t) \sigma_{ij}\, dx^i dx^j
\end{equation}
where $N$ is the lapse function, $a$ is the scale factor,  and $ \sigma_{ij}$ is the metric of the constant-curvature spatial sections. With the help of 
thermodynamical considerations, and taking the constraints into account, insertion of (\ref{FRWmetric}) into (\ref{actionSchutz})
yields  the reduced action \cite{Alvarenga} 

\begin{equation}
\label{reducedaction}
{\cal S}= \int ({\dot a}\, p_a + {\dot T}\, p_T - N\, {\cal H}) dt
\,\,\, ,
\end{equation}
with the super-Hamiltonian

\begin{equation}
\label{superHamiltonian}
{\cal H} = -\frac{p_a^2}{24a} - 6k\, a + \frac{p_T}{a^{3w}}
\,\,\, .
\end{equation}
Here $T$ is a new canonical variable \cite{Alvarenga} that describes the only remaining degree of freedom of the fluid, while $k=0,1$ or $-1$ depending 
on whether the curvature of the spatial sections is zero, positive or negative.  From now on we shall discuss only the flat case ($k=0$).

With $k=0$ the equations of motion are

\begin{equation}
\label{Hamiltonequations1}
{\dot a} = \frac{\partial}{\partial p_a}( N{\cal H})= 
-\frac{Np_a}{12a}\,\,\, ,\,\,\, {\dot p}_a = -\frac{\partial}{\partial a}( N{\cal H})= 
-\frac{Np_a^2}{24a^2} + 3w \frac{Np_T}{a^{3w+1}}
\,\,\, ,
\end{equation}

\begin{equation}
\label{Hamiltonequations2}
{\dot T} = \frac{\partial}{\partial p_T}( N{\cal H})= 
\frac{N}{a^{3w}}\,\,\, ,\,\,\, {\dot p}_T = -\frac{\partial}{\partial T}( N{\cal H})= 0 
\,\,\, ,
\end{equation}
supplemented by the super-Hamiltonian constraint

\begin{equation}
\label{superHamiltonianconstraint}
{\cal H} = -\frac{p_a^2}{24a} + \frac{p_T}{a^{3w}} = 0
\,\,\, .
\end{equation}
Solving this  constraint for $p_a$  we get

\begin{equation}
\label{constraintsolution}
p_a = - \sqrt{24p_T}\, a^{-(3w -1)/2}
\,\,\, ,
\end{equation}
where $p_T$ is a positive constant, and the negative sign has been selected in order to produce an 
expanding universe. In the cosmic-time gauge $N=1$  the first of 
equations  (\ref{Hamiltonequations1}) becomes

\begin{equation}
\label{equationscalefactor}
{\dot a} = \frac{\sqrt{24p_T}}{12}\, a^{-(3w +1)/2}
\,\,\, ,
\end{equation}
which is solved by

\begin{equation}
\label{solutionscalefactor}
a^{3(w +1)/2} = a_0^{3(w +1)/2} + \sqrt{\frac{p_T}{24}}\, 3(w +1)\, (t-t_0)
\,\,\, ,
\end{equation}
\\
where $a_0$ is the present value of the scale factor. If $\, w < -1\,$ then $\, w +1 = -\vert w +1\vert <0\,$ and 

\begin{equation}
\label{scalefactorBigRip}
a(t) = \Bigl[ a_0^{3(w +1)/2} - \sqrt{\frac{p_T}{24}}\, 3\vert w +1\vert\, (t-t_0)\Bigr]^{-2/3
\vert w +1\vert}
\,\,\, ,
\end{equation}
which becomes infinite at the  finite future time

\begin{equation}
\label{timeBigRip}
t = t_0 +  \sqrt{\frac{24}{p_T}}\,
\frac{a_0^{3(w +1)/2}}{3\vert w +1\vert}
\,\,\, .
\end{equation}
\\
This is the famous Big Rip \cite{Kamionkowski}.

\section{Quantization with a special inner product}

The Wheeler-DeWitt equation results from making the correspondence $p_a \to {\hat p}_a = -i\partial /\partial a$, $p_T \to {\hat p}_T = i\partial /\partial
 \tau$ and requiring ${\hat{\cal H}}\, \Psi =0$, where   ${\hat{\cal H}}$ is the operator corresponding to the super-Hamiltonian $\cal H$. This means
 that the time has been chosen as $\tau=-T$. With a particular choice of operator ordering the Wheeler-DeWitt equation becomes

\begin{equation}
\label{Wheeler-DeWittequation}
i\, a^{1-3w}\, \frac{\partial\Psi}{\partial \tau}= -\frac{1}{24}\,\frac{\partial^2 \Psi}{\partial a^2}
\,\,\, .
\end{equation}
As discussed in \cite{Alvarenga}, this equation takes the form of a Schr\"odinger equation $i\partial \Psi/\partial \tau = {\hat H} \Psi$ with a Hermitian 
Hamiltonian operator ${\hat H}$ if
the inner product is 

\begin{equation}
\label{innerproduct}
(\Phi,\Psi) = \int_0^{\infty} a^{1-3w}\, \Phi (a,\tau)^* \Psi (a,\tau)\, da
\,\,\, .
\end{equation}

By superposing stationary states a solution to (\ref{Wheeler-DeWittequation}) can be found \cite{Alvarenga} in the form ($\gamma$ is a positive 
constant)

\begin{equation}
\label{wavepacket1}
\Psi (a,\tau ) = a \exp \bigg\{- \frac{a^{3(1-w)}}{4[\gamma - i3(1-w)^2 \,\tau /32]} \bigg\}
\,\,\, .
\end{equation}
The expectation value of the scale factor is  \cite{Alvarenga}

\begin{equation}
\label{expectation1}
\langle a \rangle (\tau) \propto (b^2 + \tau^2)^{1/3(1-w )}\,\, , \,\, b= \frac{3(1-w)^2}{32\gamma }
\,\,\, .
\end{equation}
Note, however, that $\tau$ is not cosmic time.  Recall that $\tau =-T$, where $T$ is the only remaining degree of freedom of the fluid. It follows
 from  the first of equations (\ref{Hamiltonequations2}) that $\vert N\vert = a^{3w}$. Since $\langle a \rangle$ plays the role of the scale factor in the quantum theory, we write $dt=\langle a \rangle^{3w}\, d\tau$. Therefore, choosing a common origin for both times,

\begin{equation}
\label{cosmictime}
t\propto \int_0^{\tau} (b^2 + \lambda^2)^{w/(1-w)}\, d\lambda 
\,\,\, .
\end{equation}
Near infinity the above integral behaves as

\begin{equation}
\label{cosmictimenearinfinity}
\int^{\tau} \lambda^{2w/(1-w)} d\lambda  \propto \tau^{(1+w)/(1- w)} 
 \,\,\, .
\end{equation}
If $w<-1$ the integral is convergent and $t\to t_{rip}$ for $\tau \to \infty$, where  $t_{rip}$ is finite. Thus the expectation value of the scale factor
 becomes infinite at the finite cosmic time  $t_{rip}$ which corresponds to $\tau =\infty$. The quantum model displays a Big Rip just like the classical model.


One might suspect that this result lacks generality in view of: (i) the particular order of noncommuting operators  employed in the construction of the
 Wheeler-DeWitt 
equation (\ref{Wheeler-DeWittequation}); (ii) the unusual inner product (\ref{innerproduct}) that, oddly enough, depends on the equation of 
state through
$w$.
 We proceed to address  these issues.

\section{Operator ordering and quantization with the standard inner product}

The classical Hamiltonian corresponding to the Wheeler-DeWitt equation (\ref{Wheeler-DeWittequation}) is

\begin{equation}
\label{classicalHamiltonian}
H=  \frac{a^{3w -1}}{24}\, p_a^2
\,\,\, ,
\end{equation}
to which we associate the Hamiltonian operator

\begin{equation}
{\hat H}_{\lambda\mu\nu} = \frac{1}{24}\, \frac{1}{2} \Bigl( a^{\lambda}\, {\hat p}_a\, a^{\mu} {\hat p}_a\, 
a^{\nu} +   a^{\nu}\, {\hat p}_a\, a^{\mu} {\hat p}_a 
a^{\lambda}\Bigr)\,\,\, ,\,\,\, \lambda + \mu + \nu = 3w -1
\,\,\, .
\end{equation}
This operator is formally self-adjoint for any choice of the ordering 
parameters $\lambda$, $\mu$, $\nu$ with the standard inner product

\begin{equation}
\label{standardinnerproduct}
 (\Phi,\Psi) = \int_0^{\infty}  \Phi (a,\tau)^* \Psi (a,\tau)\, da
\,\,\, .
\end{equation}

Now the Wheeler-DeWitt equation takes the form

\begin{equation}
\label{WDWordered}
 i\frac{\partial\Psi}{\partial\tau}={\hat H}_{\lambda\mu\nu}\Psi\,\,\, ,
\end{equation}
whose stationary solutions  $\Psi(a,\tau )=\psi(a)e^{-iE\tau}$ satisfy

\begin{equation}
\label{WDWorderedstationary}
{\hat H}_{\lambda\mu\nu}\psi-E\psi=0
\,\,\, .
\end{equation}
After some algebraic labor this equation can be put in the form

\begin{equation}
\label{WDWorderedexplicit}
a^{2}\frac{d^{2}\psi}{da^{2}}+2ra\frac{d\psi}{da}+[\epsilon+24Ea^{2(1-r)}]\psi=0
\end{equation}

\noindent where $\epsilon=[\nu(2r-\lambda-1)+\lambda(2r-\nu-1)]$ and $r=(3w-1)/2$. The solutions of this equation are \cite{Hildebrand} 

\begin{equation}
\label{stationarysolutions1}
\psi (a) =a^{(1-2r)/2}Z_{p}\bigg(\frac{\sqrt{24E}}{1-r}a^{1-r}\bigg)\, ,\ \ r\neq1
\end{equation}

\noindent where 

\begin{equation}
\label{orderBesselfunction}
p=\frac{1}{1-r}\sqrt{\bigg(\frac{1-2r}{2}\bigg)^{2}-\epsilon}
\end{equation}

\noindent and $Z_{p}$ is a Bessel function  either of the first or of the second kind  
of order $p$.
We shall consider only  $Z_{p}=J_{p}\,$, the Bessel function  of the first kind, 
in which case a complete solution to equation (\ref{WDWordered}) is given by

\begin{equation}
\label{WDWstationarysolution}
\Psi_E(a,\tau )= A\, e^{-iE\tau}\,a^{(1-2r)/2}\, J_{p}\bigg(\frac{\sqrt{24E}}{1-r}a^{1-r}\bigg).
\end{equation}
Our next step is to generate a normalizable wave packet by superposition  of the solutions  (\ref{WDWstationarysolution}) in the form

\begin{equation}
\label{superposition1}
\Psi(a,\tau )=\int_{0}^{\infty}dEA(E)e^{-iE\tau}a^{(1-2r)}J_{p}\bigg(\frac{\sqrt{24E}}{1-r}a^{1-r}\bigg)
\,\,\, .
\end{equation}
Setting $\lambda =\sqrt{24E}/(1-r)$ this takes the more convenient form

\begin{equation}
\label{superposition2}
\Psi(a,\tau )=\int_{0}^{\infty}d\lambda C(\lambda)e^{-i(1-r)^{2}\lambda^{2}\tau/24}a^{(1-2r)}J_{p}(\lambda a^{1-r})\,\,\, .
\end{equation}
The choice

\begin{equation}
\label{superpositionchoice}
C(\lambda)=\lambda^{p+1}e^{-\gamma\lambda^{2}}\,\,\, ,\,\,\, \gamma>0
\end{equation}
allows the integration to be performed in closed form and we obtain \cite{Gradshteyn}

\begin{equation}
\label{wavepacketarbitraryordering}
\Psi(a,\tau )=a^{-1/2}\bigg[\frac{a^{1-r}}{2\gamma+\frac{i(1-r)^{2}\tau}{12}}\bigg]^{1+p}\exp\bigg[-\frac{a^{2(1-r)}}
{4\gamma+\frac{i(1-r)^{2}\tau}{6}}\bigg]
\,\,\, .
\end{equation}

The expectation value of the scale factor is defined by

\begin{equation}
\label{expectationdefinitionarbitraryordering}
\langle a \rangle (\tau) = \frac{\int_0^{\infty} a\, \vert\Psi (a,\tau )\vert^2\, da}{\int_0^{\infty} \vert\Psi (a,\tau )\vert^2\, da}
\end{equation}
and a  straightforward computation furnishes

\begin{equation}
\label{expectation2}
\langle a \rangle (\tau)=\frac{\Gamma\bigg[\frac{2(1-r)(1+p)+1}{2(1-r)}\bigg]}{\Gamma (p+1)}\bigg\{\frac{1}{2\gamma}
\bigg[\gamma^{2}+\frac{(1-r)^{4}}{576}\tau^{2}\bigg]\bigg\}^{1/2(1-r)}
\,\,\, .
\end{equation}
Recalling that $r=(3w-1)/2$,  equation (\ref{expectation2}) yields

\begin{equation}
\label{expectation2coincident}
\langle a \rangle (\tau)\propto (b^2 + \tau^2)^{1/3(1-w)}
\,\,\, ,
\end{equation}
the same as (\ref{expectation1}). Thus, an arbitrary operator order with the standard inner product leads to the same result as the one obtained when  a special operator order is
 adopted which requires the strange inner product
(\ref{innerproduct}). This lends support to our previous claim that there exist evolving  quantum states such  that the  phantom energy  gives rise 
 to accelerated expansion and a Big Rip.

It should be stressed that we are trying to find out whether quantum effects necessarily remove the future singularity. If this were the case,
for {\it any} quantum state the expectation value of the scale factor would have to remain finite for any  value of the cosmic time. We 
have just presented  an example of a quantum state such that  the expectation value of the scale factor 
becomes infinite in a finite cosmic time. This shows that, at least in this model, and for the above quantization procedure, it is not generally the case that 
quantum effects prevent  the Big Rip. This reasoning also justifies our particular choices (\ref{WDWstationarysolution}) and (\ref{superpositionchoice}), 
which allow the integration to be performed  and the wave packet to be expressed in terms of elementary functions, which, in turn, makes easy the 
computation of the expectation value of the scale factor.

\section{Canonical transformation and another quantization with the standard inner product}

The phase-space transformation $(a,p_a )\to (x,p)$ defined by

\begin{equation}
\label{canonicaltransformation}
x= \frac{2\sqrt{12}}{3(1-w)}\, a^{3(1-w)/2}\,\,\, ,\,\,\, p = \frac{a^{(3w -1)/2}}{\sqrt{12}}\, p_a
\end{equation}
is canonical inasmuch as the Poisson bracket $\{x,p\}_{(a,p_a)}=1$. In terms of these new canonical variables the classical
Hamiltonian (\ref{classicalHamiltonian}) reduces to
the free-particle Hamiltonian

\begin{equation}
\label{free-particleHamiltonian}
H= \frac{p^2}{2}
\,\,\, ,
\end{equation}
and the corresponding Wheeler-DeWitt equation takes the form 

\begin{equation}
\label{free-particleWheeler-DeWitt}
i\,  \frac{\partial\Psi}{\partial \tau}= -\frac{1}{2}\,\frac{\partial^2 \Psi}{\partial x^2}
\,\,\, .
\end{equation}
This is the Schr\"odinger wave equation for a free particle on the half-line $[0,\infty )$. Self-adjointness of the Hamiltonian operator requires \cite{Reed}
 that the 
the domain of  ${\hat H}$ be restricted to those wave functions that satisfy the boundary
condition

\begin{equation}
\label{boudarycondition}
\Psi (0,\tau) = \xi\,\Psi^{\prime} (0,\tau)
\end{equation}
where $\Psi^{\prime}=\partial \Psi/\partial x$ and 
 $\xi\in [0,\infty ]$.
For the sake of simplicity we shall consider only the cases $\xi =0$
and $\xi =\infty$, that is, either the wave function or its derivative vanishes at $x=0$. The propagator for arbitrary $\xi$ is known  \cite{Clark}, but it is 
sufficiently 
complicated to 
prevent us from constructing  manageable  wave packets. However, inasmuch as we are searching  for examples of quantum  states that lead to  a
 future singularity,
if we are successful our particular choices of $\xi$ will still allow us to conclude that it is not true that the Big Rip is   suppressed {\it in general}, that is, for arbitrary $\xi$. 
In short, our special choices for $\xi$ are technically convenient and do not impair the logic of our reasoning. 


The propagators for the cases $\xi =0$ and $\xi =\infty$
 are given by \cite{Clark}

\begin{equation}
\label{propagators}
G_0 (x,x^{\prime},\tau ) = G(x,x^{\prime},\tau ) - G(x,-x^{\prime},\tau ) \,\,\, ,\,\,\,
G_{\infty} (x,x^{\prime},\tau ) = G(x,x^{\prime},\tau ) + G(x,-x^{\prime},\tau )
 \,\,\, ,
\end{equation}
where $G(x,x^{\prime},\tau )$ is the usual free-particle propagator, which in our case ($m=1$, $\hbar=1$) takes the form \cite{Schiff}

\begin{equation}
\label{usualpropagator}
G(x,x^{\prime},\tau ) =\bigg(\frac{1}{2\pi i \tau}\bigg)^{1/2}\exp \bigg[i\,\frac{(x-x^{\prime})^2}{2\tau} \bigg]
 \,\,\, .
\end{equation}

Let us first consider the initial normalized wave function

\begin{equation}
\label{initialwavefunctioneven}
\Psi_e (x,0)= \bigg(\frac{8\sigma}{\pi}\bigg)^{1/4} e^{-\beta x^2}\,\,\, ,\,\,\, \beta = \sigma +i \gamma\,\,\, ,\,\,\, \sigma >0
 \,\,\, .
\end{equation}
Since its derivative vanishes at $x=0$, it satisfies the boundary condition 
(\ref{boudarycondition}) with
 $\xi =\infty$. Therefore

\begin{equation}
\label{wavefunctioneven}
\Psi_e (x,\tau)= \int_0^{\infty} G_{\infty} (x,y,\tau)\Psi_e (y,0)\, dy =
\int_{-\infty}^{\infty} G(x,y,\tau)\Psi (y,0)\, dy 
\,\,\, ,
\end{equation}
where we have taken advantage of the fact that $\Psi_e (x,0)$ is an even function to extend the integration to the whole real line. A simple 
computation yields

\begin{equation}
\label{wavepacketeven}
\Psi_e (x,\tau)=  \bigg(\frac{8\sigma}{\pi}\bigg)^{1/4}(1+2i\beta \tau )^{-1/2}\exp\bigg[-\frac{\beta \,x^2}{1+2i\beta\tau}\bigg]
\,\,\, .
\end{equation}
In accordance with (\ref{canonicaltransformation}) the expectation value of the scale factor is defined by 

\begin{equation}
\label{expectationdefinitioncanonical}
\langle a \rangle_e (\tau) = \bigg[\frac{3(1-w )}{2\sqrt{12}}\bigg]^{2/ 3(1-w)} \int_0^{\infty} x^{2/ 3(1-w)}
\, \vert\Psi_e(x,\tau )\vert^2\, dx
\,\,\, ,
\end{equation}
and a simple calculation gives 

\begin{equation}
\label{expectationeven}
\langle a \rangle_e (\tau)=\frac{1}{\sqrt{\pi}}\,\bigg[\frac{3(1-w)}{2\sqrt{12}}\bigg]^{2/3(1-w)}\, \Gamma\bigg[\frac{5-3w}{3(1-w )}\bigg]
\bigg[\frac{4\sigma^2 \tau^2 + (1-2\gamma\tau )^2}{2\sigma}\bigg]^{1/3(1-w )}
\,\,\, ,
\end{equation}
whose asymptotic behavior coincides with that of (\ref{expectation1}).

Finally, consider the initial normalized wave function

\begin{equation}
\label{initialwavefunctionodd}
\Psi_o (x,0)= \bigg(\frac{128\sigma^3}{\pi}\bigg)^{1/4}\, x\,  e^{-\beta x^2}\,\,\, ,\,\,\, \beta = \sigma +i \gamma\,\,\, ,\,\,\, \sigma >0
\,\,\, ,
\end{equation}
which satisfies  the boundary condition 
(\ref{boudarycondition}) with
 $\xi =0$.  As in the previous case, the fact that $\Psi_o (x,0)$ is an odd function allows us to write

\begin{equation}
\label{wavefunctionodd}
\Psi_o (x,\tau)= \int_0^{\infty} G_{0} (x,y,\tau)\Psi_o (y,0)\, dy =
\int_{-\infty}^{\infty} G(x,y,\tau)\Psi_o (y,0)\, dy 
\,\,\, ,
\end{equation}
which gives

\begin{equation}
\label{wavepacketodd}
\Psi_o (x,\tau)=  \bigg(\frac{128\sigma^3}{\pi}\bigg)^{1/4}(1+2i\beta \tau )^{-3/2}
\, x\,\exp\bigg[-\frac{\beta \,x^2}{1+2i\beta\tau}\bigg]
\,\,\, .
\end{equation}
The expectation value of the scale factor is now given by

\begin{equation}
\label{expectatiodd}
\langle a \rangle_o (\tau)= \frac{5-3w}{3(1-w )\sqrt{\pi}}\,\bigg[\frac{3(1-w)}{2\sqrt{12}}\bigg]^{2/3(1-w)}\,
 \Gamma\bigg[\frac{5-3w}{3(1-w )}\bigg]
\bigg[\frac{4\sigma^2 \tau^2 + (1-2\gamma\tau )^2}{2\sigma}\bigg]^{1/3(1-w )}
\,\,\, ,
\end{equation}
or, equivalently,

\begin{equation}
\label{expectationevenodd}
\langle a \rangle_o (\tau)= \frac{5-3w}{3(1-w )}\, \langle a \rangle_e (\tau)
\,\,\, .
\end{equation}
Thus, we recover the same asymptotic behavior of the previous case,  which entails  accelerated expansion and a Big Rip.

\section{Conclusions}

We have investigated the quantum features of a FRW cosmological model dominated by a phantom energy fluid. The classical model was formulated on the 
basis of the canonical formalism of Schutz, which takes into account the  degrees of freedom of the fluid. In the case of a phantom fluid (equation of 
state $p=w \rho$ with $w < -1$) the classical equations of motion predict an accelerated expansion for the Universe ending in a catastrophic Big Rip a
 finite time from now. Then the model was quantized in three different ways. Firstly, the Wheeler-DeWitt equation was set up with a special choice of the 
order of noncommuting operators, which requires an unusual inner product that depends on the equation of state for the fluid. It was found that the  
expectation value 
of the scale factor becomes infinite at a finite cosmic time, and the Universe winds up in a  Big Rip. Next, an arbitrary order of  
noncommuting operators was allowed for the purpose
of quantizing the model with the standard inner product. A particular wave packet was constructed, and it was found that the  expectation value 
of the scale factor reproduces the previous behavior. Finally, a third method of quantization was explored based on the previous reduction of the 
classical Hamiltonian to that of a free particle by means of a canonical transformation. Since the propagator of the Schr\"odinger
equation for a free particle on the half-line is  known, it is possible, in principle, to find explicitly the wave function at any time once  its form is given at 
an initial time. It turns out that two initial wave functions obeying different boundary conditions give rise to essentially the same expectation value 
of the scale factor with exactly the same behavior predicted by the two other quantization schemes. This indicates that our results are not  merely an
 artifact of a 
spurious quantization procedure. The  quantum behavior of the scale
factor appears to be robust, and supports our conclusion that, at least in our model,  quantum effects do not, in general,  prevent  a Big Rip. 

We do not claim that 
our result is conclusive.
 In order to check its generality,  
the matter content should be enriched to  describe the early universe more realistically, which we intend
to do in a future work. Since the phantom energy dominates the late time dynamics, we suspect that  the enrichment of the matter 
content will not change our result, but this conjecture must be checked. We anticipate that this will not be an easy task
because it will surely be  much more difficult to obtain exact normalizable solutions to
the Wheeler-DeWitt equation. We do not expect that approximate solutions,
such as those furnished by the WKB method, are suitable to discuss the existence of
quantum singularities, that is, we believe that  only a fully quantum treatment can give us a solid  clue as to the persistence
of future singularities in  the quantum regime of phantom cosmological models.
We also feel that  a study of the 
Bohmian trajectories of the scale factor is likely to be helpful in clarifying  whether quantum gravitational
 effects can suppress the Big Rip. 
This is an investigation we expect to undertake in the near future. 


Our result differs from those obtained by  means of  other formalisms \cite{Odintsov,Sami,Kiefer}, namely that  the Big Rip is 
softened or  avoided by quantum effects. On the one hand, since it is not known how these different formulations of quantum cosmology are related, 
we confess to our ignorance about  
the reason for the discrepancy. On the other hand,  it is hard 
 to make clear-cut statements on presence or absence of cosmological  singularities  in the quantum regime because there is no general 
agreement on  what constitutes a quantum singularity. 
 These issues certainly deserve 
a deeper investigation.

\begin{acknowledgments}
The work of Ed\'esio M. Barboza Jr. was partially supported by Coordena\c c\~ao de Aperfei\c coamento de Pessoal de N\'{\i}vel Superior
(CAPES), Brazil.
\end{acknowledgments}

\end{document}